\documentclass[sigconf,nonacm]{acmart}

\usepackage{graphicx}  
\usepackage{booktabs}  
\usepackage{amsmath}   
\usepackage{hyperref}  
\usepackage{multirow}  
\usepackage{balance}   
\usepackage[english]{babel}
\usepackage{microtype}  
\usepackage{tabularx}
\usepackage[justification=centering]{caption}
\usepackage{subcaption}  
\usepackage{enumitem}

\AtBeginDocument{%
  }

\makeatletter
\DeclareRobustCommand\onedot{\futurelet\@let@token\bmv@onedotaux}
\def\bmv@onedotaux{\ifx\@let@token.\else.\null\fi\xspace}
\def\eg{\emph{e.g}.} 
\def\ie{\emph{i.e}.} 
 
 \def\vs{\emph{vs}.}

\makeatother

\begin{document}

\title{Truth and Trust: Fake News Detection via Biosignals
}

\pagestyle{plain}


\author{Gennie Nguyen}
\affiliation{%
  \institution{The Australian National University}
  \city{Canberra}
  \state{ACT}
  \country{Australia}
}
\email{Gennie.Nguyen@anu.edu.au}

\author{Lei Wang}
\affiliation{%
  \institution{Griffith University}
  \city{Brisbane}
  \state{Queensland}
  \country{Australia}
}
\email{l.wang4@griffith.edu.au}

\author{Yangxueqing Jiang}
\affiliation{%
  \institution{School of Medicine and Psychology, The Australian National University}
  \city{Canberra}
  \state{ACT}
  \country{Australia}
}
\email{Yangxueqing.jiang@anu.edu.au}

\author{Tom Gedeon}
\affiliation{%
  \institution{Optus Centre for AI, Curtin University}
  \city{Perth}
  \state{Western Australia}
  \country{Australia}
}
\email{tom.gedeon@curtin.edu.au}


\begin{abstract}
Understanding how individuals physiologically respond to false information is crucial for advancing misinformation detection systems. This study explores the potential of using physiological signals, specifically electrodermal activity (EDA) and photoplethysmography (PPG), to classify both the veracity of information and its interaction with user belief. In a controlled laboratory experiment, we collected EDA and PPG signals while participants evaluated the truthfulness of climate-related claims. Each trial was labeled based on the objective truth of the claim and the participant's belief, enabling two classification tasks: binary veracity detection and a novel four-class joint belief-veracity classification. We extracted handcrafted features from the raw signals and trained several machine learning models to benchmark the dataset. Our results show that EDA outperforms PPG, indicating its greater sensitivity to physiological responses related to truth perception. However, performance significantly drops in the joint belief-veracity classification task, highlighting the complexity of modeling the interaction between belief and truth. These findings suggest that while physiological signals can reflect basic truth perception, accurately modeling the intricate relationships between belief and veracity remains a significant challenge. This study emphasizes the importance of multimodal approaches that incorporate psychological, physiological, and cognitive factors to improve fake news detection systems. Our work provides a foundation for future research aimed at enhancing misinformation detection via addressing the complexities of human belief and truth processing.

 
\end{abstract}




\maketitle

\section{Introduction}

\begin{figure*}[h]
\centering
\begin{subfigure}[t]{0.38\textwidth}
    \centering
    \includegraphics[width=0.8\linewidth]{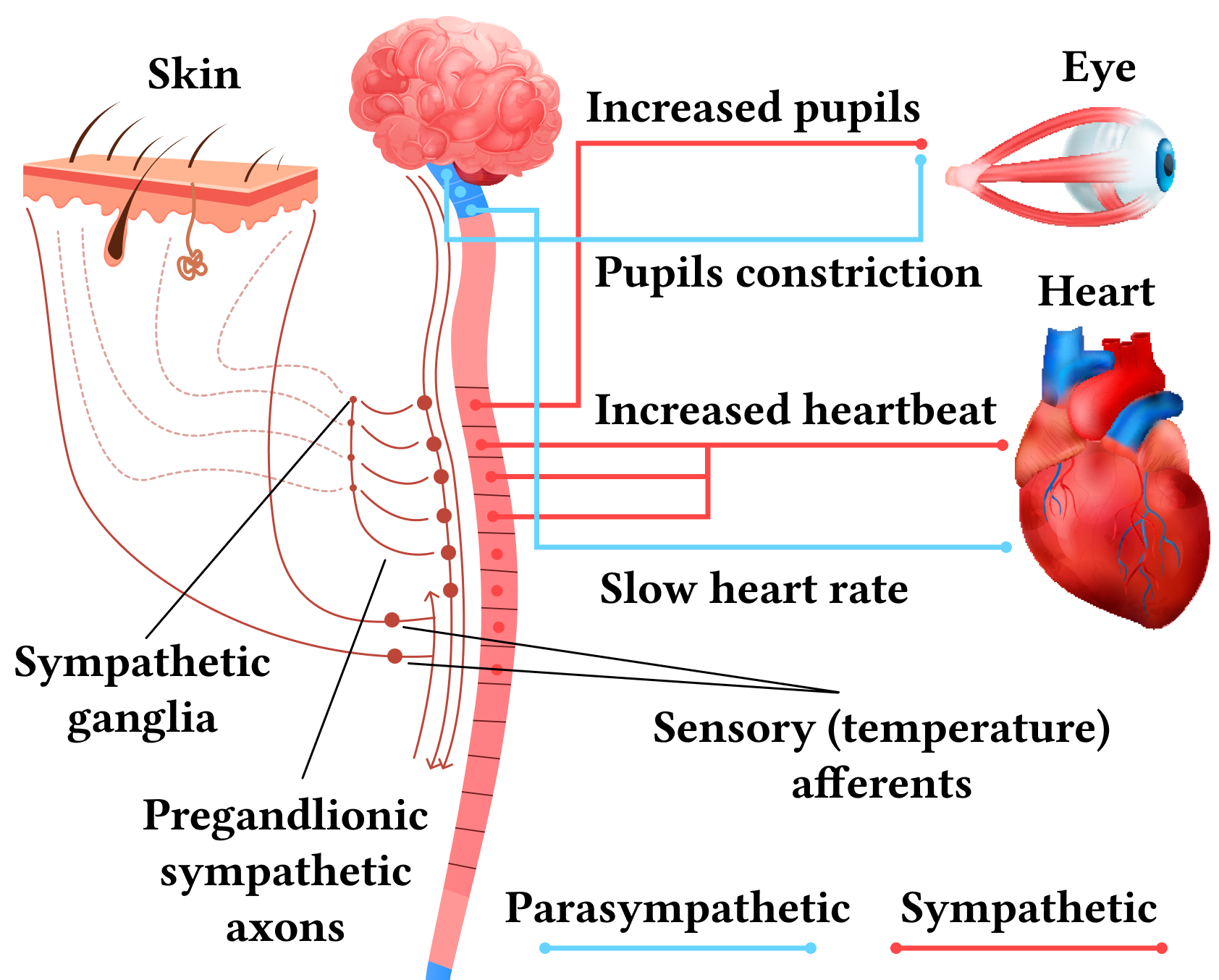}
    \caption{\textbf{Autonomic nervous system.} The sympathetic and parasympathetic branches regulate bodily responses involved in arousal, attention, and cognitive conflict.}
    \label{fig:ans}
\end{subfigure}
\hfill
\begin{subfigure}[t]{0.30\textwidth}
    \centering
    \includegraphics[width=\linewidth]{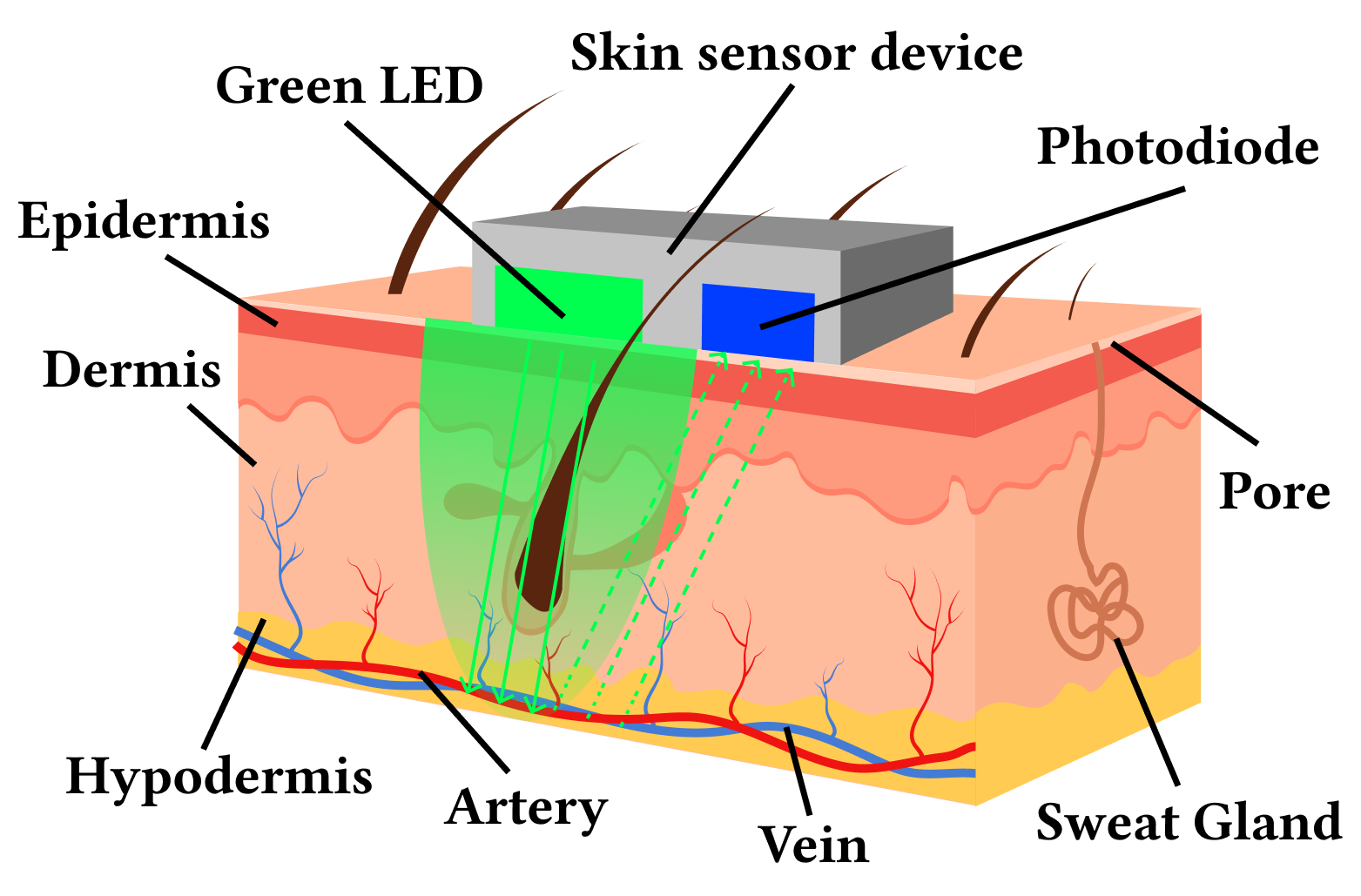}
    \caption{\textbf{Sensor detection.} Skin-based sensors capture changes in sweat conductance and blood volume to infer internal states.}
    \label{fig:skin_detection}
\end{subfigure}
\hfill
\begin{subfigure}[t]{0.30\textwidth}
    \centering
    \includegraphics[width=0.9\linewidth]{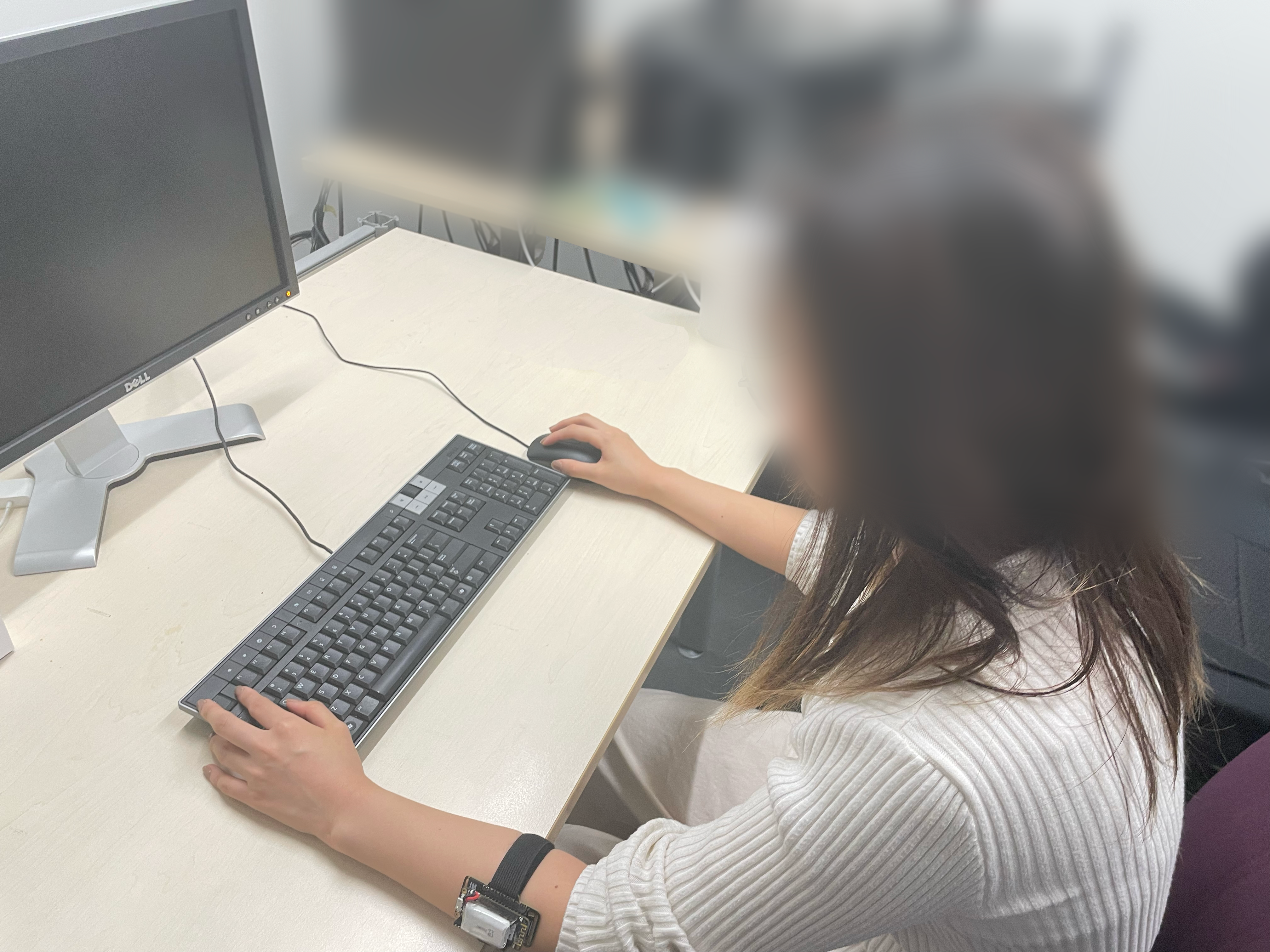}
    \caption{\textbf{Participant during recording.} EDA and PPG were captured as participants judged climate-related claims.}
    \label{fig:participant}
\end{subfigure}

\caption{
\textbf{Physiological foundations for detecting responses to misinformation.}  
Our study investigates whether internal physiological signals reflect how people process (mis)information.  
(a) The autonomic nervous system governs involuntary responses, such as heart rate and sweat gland activity, that accompany cognitive and emotional states.  
(b) These responses are measured non-invasively via electrodermal activity (EDA) and photoplethysmography (PPG), which reflect arousal, stress, and attention.  
(c) During the experiment, participants wore an EmotiBit sensor that continuously recorded EDA and PPG while they evaluated the veracity of climate-related claims.  
By mapping these physiological signals to belief and truth judgments, our work explores how bodily responses can inform human-centered misinformation detection systems.
}
\label{fig:physio_overview}
\end{figure*}

The widespread dissemination of misinformation across digital platforms has created an urgent need for systems that can detect and mitigate its influence. 
While computational methods have made significant strides by analyzing linguistic features \cite{fang2024signllm, chen2024sato}, content style \cite{wang2024flow, wang2024high}, and source credibility \cite{zhuadvancing}, these approaches largely overlook how misinformation is experienced by individuals. 
Yet, understanding the human dimension, how people physiologically and cognitively respond to false information, may offer critical insights into the detection and design of more robust defense systems \cite{wang2023robust,10.1145/3701716.3717744}.

Physiological signals offer a promising but underexplored avenue in this space. Signal types such as electrodermal activity (EDA), photoplethysmography (PPG), and EEG have long been used to study deception, emotional arousal, and cognitive conflict in areas like lie detection and affective computing \cite{ulmandevelopment, speth2021deception, joshi2025multimodal, pandey2019techniques, rahal2024your, van2015unconscious, gunderson2023body, chu2021detecting}. However, much of this research focuses on individuals who are themselves deceiving others, or on interpersonal deception where intent and awareness are central. These studies are typically situated in high-stakes, face-to-face contexts and aim to uncover cues associated with intentional dishonesty (see Fig.~\ref{fig:physio_overview} (a) and (b)).

In contrast, misinformation in digital environments is often impersonal, diffuse, and unintentional. A false claim encountered online may not involve an identifiable deceiver, yet it can still shape beliefs and behaviors. 
This shift calls for new paradigms of analysis: rather than studying deception as a performed act, we investigate how receiving false information impacts the human body \cite{dingjourney, wang2025time}.
Only a handful of studies have begun to probe this space, their results suggesting that users may exhibit physiological responses when exposed to misleading or incongruent content \cite{rahal2024your, van2015unconscious, gunderson2023body, chu2021detecting}. However, these works often lack fine-grained annotations of belief or veracity, and few systematically explore how objective truth and subjective belief interact at the physiological level \cite{zhu2018detecting, ten2019different}.

This study addresses that gap. In a controlled laboratory setting, we collected EDA and PPG signals as participants evaluated the veracity of climate-related claims (see Fig.~\ref{fig:physio_overview} (c)). Each trial was annotated with both the objective truth of the statement (True or False) and the participant’s subjective belief (Believe or Not Believe). This dual labeling allows us to explore two complementary tasks: (i) \textit{Binary veracity classification}: Can physiological signals distinguish between objectively true and false information, regardless of user belief? (ii) \textit{Joint belief-veracity classification}: Do physiological responses capture more nuanced interactions, such as when someone believes a false claim or correctly rejects a true one?

By analyzing these tasks, we investigate whether low-cost, non-invasive biosignals can reveal latent traces of how truth and belief are processed. 
Our findings indicate that EDA, in particular, shows greater sensitivity to truth-related arousal than PPG. 
However, classification performance drops significantly in the joint task, underscoring the complexity of modeling how truth and belief interact in the body.

Beyond technical performance, our work contributes conceptually to the design of future misinformation detection systems that are not purely content-driven, but also responsive to how content is experienced. As misinformation becomes more pervasive and personalized, such human-centered, multimodal approaches are essential \cite{wang2017analysis, wang2024flow, wang2024high, ding2025learnable}. In summary, this paper contributes:
\renewcommand{\labelenumi}{\roman{enumi}.}
\begin{enumerate}[leftmargin=0.6cm]
\item \textbf{A novel dataset} combining EDA, PPG, video recordings, and user self-reports, annotated with both belief and veracity labels for each trial.
\item \textbf{Two classification tasks} that isolate physiological responses to objective truth and explore the interaction between belief and veracity. 
\item \textbf{Comparative analysis} of EDA and PPG, demonstrating EDA’s superior sensitivity in detecting veracity-related physiological cues.
\item \textbf{Insight} into the psychophysiological complexity of misinformation processing, revealing challenges in modeling nuanced cognitive-affective states from biosignals.
\end{enumerate}

Together, these contributions offer a new perspective on how physiological signals can illuminate the subtle interplay between what we read, what we believe, and what is true.

\section{Related work}

Prior research on misinformation and veracity detection has predominantly focused on content-based methods, using linguistic features, semantic coherence, or patterns in social network propagation to differentiate true from false claims \cite{datta2024enhancing, bhattacharjee2017active, kumar2019rumour}. While these approaches have shown promise, they largely overlook the underlying physiological and cognitive processes that influence how people perceive, evaluate, and respond to information.

In parallel, the fields of affective computing and deception detection have explored how physiological signals, such as EDA, PPG, and electrocardiography (ECG), reflect internal states like arousal, stress, and emotional valence \cite{babaei2021critique, gasparini2021deep, kang20221d, lin2023review}. These studies suggest that bodily responses encode meaningful cues related to human judgment, particularly in contexts involving uncertainty, belief, or deception.

\begin{figure*}[tbp]
  \includegraphics[width=\textwidth]{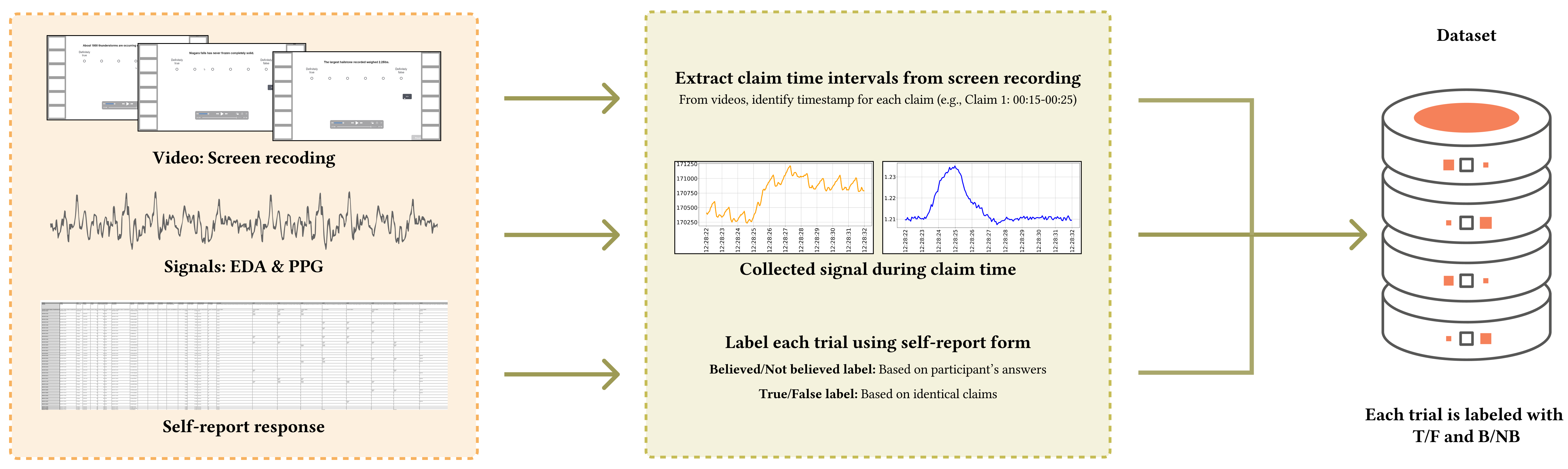}
  \caption{
  Overview of the dataset construction process used for veracity and belief classification tasks. The dataset comprises three modalities: (i) screen recordings of participants evaluating climate-related claims (video), (ii) physiological signals: electrodermal activity (EDA) and photoplethysmography (PPG), recorded in CSV format, and (iii) participants’ self-reported belief responses (CSV). Timestamps of each claim presentation are extracted from the videos to isolate relevant time windows, which are then used to align the corresponding physiological data. Each trial is labeled with both a belief label (Believe or Not Believe) and a veracity label (True or False), enabling two downstream tasks: binary veracity classification and four-class joint belief-veracity classification. This multimodal, time-aligned dataset supports analysis of physiological responses to misinformation and the complex interplay between belief and truth.
  }
  \label{fig:multi-modalities}
\end{figure*}

\textbf{Physiological signals in deception detection.} Physiological signals have long played a central role in deception detection \cite{national2003polygraph, wahabassessing, synnott2016review, ulmandevelopment, speth2021deception, joshi2025multimodal, pandey2019techniques}. Traditional systems \cite{national2003polygraph, hartwig2014lie, bhaskaran2011lie, horvath1982detecting, talaat2024explainable}, such as polygraphs \cite{national2003polygraph, wahabassessing, synnott2016review}, use biosignals like EDA, ECG, and respiration to infer deception by identifying physiological changes associated with lying. For example, the system developed by \citet{ulmandevelopment} exemplifies this approach, combining multiple biosignals in a real-time credibility assessment platform optimized for use in practical settings.

Recent advances have expanded upon these foundations by integrating machine learning and multimodal sensing \cite{fang2024signllm,chen2024sato,dingjourney,10.1145/3701716.3717744,ding2025learnable,wang2025time}. The DDPM dataset introduced by \citet{speth2021deception} established a benchmark for remote deception detection using signals such as EDA and skin temperature, collected in a controlled ``mock crime'' scenario. These signals were analyzed using standard machine learning models, providing early evidence that non-invasive physiological data can support automated lie detection.

A more ambitious effort is represented by the CogniModal-D dataset of \citet{joshi2025multimodal}, which captures seven modalities, including EEG, ECG, EOG, GSR, gaze, audio, and video, from over 100 participants during tasks involving interpersonal deception and emotionally charged storytelling. While neurophysiological signals (\eg, EEG, ECG) contributed to detection, behavioral signals, particularly GSR (EDA), gaze, and facial expressions, proved more effective. This study highlights the strength of combining verbal, nonverbal, and physiological cues to build more robust deception detection systems.

Complementing these empirical contributions, \citet{pandey2019techniques} conducted a systematic review of behavioral deception detection supported by physiological data. They identified EDA as a particularly reliable signal due to its involuntary nature and strong correlation with emotional arousal. Their review also cataloged common machine learning techniques in this domain, such as support vector machines, k-nearest neighbors, and discriminant analysis.

Together, these efforts illustrate the growing sophistication of physiological deception detection, from traditional polygraphs \cite{national2003polygraph} to AI-enhanced \cite{wang2023robust, wang2024meet}, multimodal systems \cite{wang2017analysis, wang2023robust, wang2024flow, wang2024high, ding2025learnable, wangtaylor,wang2024meet}. However, the majority of these studies focus on analyzing the physiological responses of deceivers, the individuals producing false information, while largely neglecting the physiological responses of observers, \ie, the recipients of deceptive content.

\textbf{Physiological signals in observer-based lie detection.} An emerging body of work investigates whether the physiological responses of observers, those evaluating deceptive content, can serve as indicators of deception. \citet{chu2021detecting} found that EDA signals recorded from participants watching political statements significantly correlated with the objective falsehood of those statements. Their deep learning models (\eg, ResNet, VAE-LSTM) achieved high F1-scores in classifying deception levels based solely on observer physiology.

\citet{gunderson2023body} further examined how interoceptive accuracy, the ability to sense internal bodily states, modulates physiological reactions to deception. They observed that individuals with higher interoceptive sensitivity exhibited stronger vasoconstriction responses when exposed to deceptive content. Interestingly, this physiological sensitivity did not improve conscious deception detection, revealing a disconnect between subconscious bodily responses and explicit judgments.

Other studies \cite{rahal2024your, van2015unconscious} have yielded more mixed findings. \citet{rahal2024your}, for example, used thermal imaging to track finger skin temperature and found no significant differences between responses to truthful and deceptive statements. Moreover, subjective ratings of trust and likability failed to distinguish liars from truth-tellers. In contrast, \citet{van2015unconscious} reported finger temperature drops in participants who were primed to suspect deception, suggesting that physiological responses may only emerge under conditions of heightened cognitive vigilance. Collectively, these findings indicate that observer-based physiological cues hold potential for lie detection, but their reliability is highly dependent on attentional focus, bodily awareness, and experimental context.

\textbf{Physiological responses to misinformation exposure.} Understanding how people physiologically respond to misinformation is crucial for designing detection systems and interventions that go beyond content analysis \cite{molina2025exploring, wang2025bridging}. \citet{deseeing} examined user responses to real and fake news articles using EDA, PPG, and eye-tracking in conjunction with sentiment analysis. They found that fake news consistently elicited higher physiological arousal, especially in EDA amplitude and heart rate variability, even though participants often failed to recognize the content as false and sometimes expressed willingness to share it. This suggests that misinformation can evoke subconscious physiological responses, even in the absence of explicit awareness.

\citet{chu2021detecting} similarly showed that observers’ EDA responses to political statements correlated with the objective degree of falsehood. Their models were able to classify statement veracity using only observer physiological data, further demonstrating that deceptive content leaves a measurable physiological footprint. Notably, they highlight two key directions for future work: (i) evaluating detection in scenarios where participants are unaware deception may occur, and (ii) investigating how cognitive factors, such as political bias or familiarity, influence physiological responses. These questions are highly relevant to real-world misinformation, where exposure often occurs passively and through the filter of personal beliefs.

In another study, \citet{joshi2025multimodal} used a ``best friend scenario'' where participants listened to emotionally charged but deceptive or truthful narratives. While EEG and audio features were relatively ineffective, behavioral and physiological signals, such as gaze and GSR, successfully distinguished deceptive exposure. These results further support the idea that misinformation affects the body at a largely unconscious level and that physiological and behavioral cues can serve as subtle, yet reliable, indicators of deceptive content.

\textbf{Gaps and contributions} Despite these advances, critical gaps remain. Most prior work has focused either on interpersonal deception detection \cite{speth2021deception, rahal2024your, van2015unconscious} or on constrained experimental analyses of physiological responses to misinformation \cite{9851990, 10705859, volz2017psychophysiological}. Few studies \cite{chu2021detecting} have attempted to systematically model the veracity of claims independent of individual belief. Even fewer have examined how belief and veracity interact to shape physiological responses.

Our work addresses these gaps by introducing two novel classification tasks: veracity detection and joint belief-veracity modeling, based solely on EDA and PPG signals collected during a controlled claim evaluation experiment. In contrast to prior approaches that rely on textual features, self-reports, or linguistic cues, our method focuses purely on physiological signals. This allows for a more content-agnostic and generalizable framework for understanding how people internalize, or reject, true and false information.

\section{Dataset and Study Design}

This work investigates whether physiological responses, specifically EDA and PPG, can reveal meaningful distinctions between true and false information, and whether these responses are modulated by a person’s belief in the information. To that end, we designed a controlled laboratory experiment, developed a novel multimodal dataset, and benchmarked a set of classification models on two tasks: binary veracity detection and a more cognitively complex joint belief-veracity classification.

\subsection{Dataset}

\textbf{Experimental protocol and labeling.}
Participants were presented with a series of short, factual climate-related claims and asked to evaluate their truthfulness while their physiological signals were recorded. The study followed a three-phase protocol \cite{jiang2024repetition}: encoding, distraction, and evaluation, adapted from prior work in truth processing and misinformation research.

In the encoding phase, participants viewed a randomized sequence of true and false claims (validated by domain experts), with each item displayed for 8 seconds. Following a 15-minute filler task to mitigate recency effects, participants entered the evaluation phase, where they re-encountered a subset of claims and judged their veracity on a 6-point Likert scale (1 denotes definitely false, 6 denotes definitely true). EDA and PPG signals were recorded throughout.

\begin{figure}[tbp]
\centering
\begin{subfigure}[t]{0.35\linewidth}
    \centering
    \includegraphics[height=5cm]{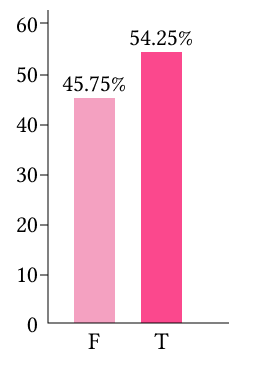}
    \caption{}
    \label{fig:truth_labels}
\end{subfigure}
\begin{subfigure}[t]{0.5\linewidth}
    \centering
    \includegraphics[height=5cm]{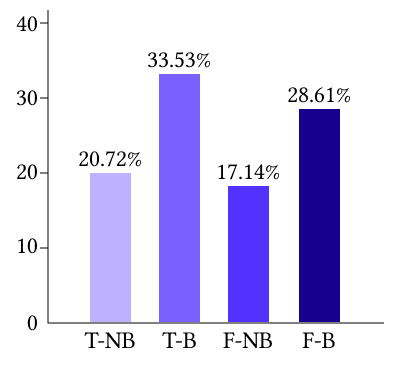}
    \caption{}
    \label{fig:four_distinct}
\end{subfigure}
\caption{Distribution of labels in our dataset.  
(a) Distribution of claim veracity labels: True (T) claims slightly outnumber False (F) claims.  
(b) Distribution across four combined belief-veracity conditions: T-NB (True–Not Believe), T-B (True–Believe), F-NB (False–Not Believe), and F-B (False–Believe).}
\label{fig:label_dist}
\end{figure}

Each trial was labeled with two complementary outcomes. First, an objective truth label (True or False), based on fact-checking. Second, a subjective belief label (Believe or Not Believe), derived from the participant’s Likert response (thresholded at the midpoint). This dual labeling enabled two classification tasks: (i) \textit{veracity classification}: discriminating true from false claims regardless of belief, and (2) \textit{joint belief-veracity classification}: capturing interactions such as believing a falsehood or disbelieving a truth. Figure \ref{fig:label_dist} shows the label distribution of our dataset.

The dataset comprises 28 participants (ages 18–34, with average age of 23.52, standard deviation of 3.28), with diverse linguistic and cultural backgrounds. Each participant completed 24 trials, yielding a total of 672 labeled examples per signal type.

\textbf{Multimodal data collection.}
In addition to physiological recordings, each session captured a screen recording video and a self-report survey. While the primary focus of this paper is on EDA and PPG analysis, the broader dataset includes rich contextual information to support future multimodal studies (see Fig.~\ref{fig:multi-modalities}).

Screen recordings provided detailed behavioral cues, such as time spent per claim, navigation patterns, and the number of changes before finalizing a response. While not modeled in this study, we used these recordings to align timestamps for trial segmentation. Self-reports recorded participants’ judgments and other metadata, including prior exposure and perceived difficulty.

The physiological signals were captured using EmotiBit devices worn on the participants’ upper arm (see Fig.~\ref{fig:physio_overview} (c)). EDA reflects sympathetic nervous system activity linked to arousal, attention, and cognitive effort. PPG, meanwhile, captures blood volume changes associated with cardiovascular regulation and cognitive load. For analysis, we retained only the segments corresponding to active claim interaction, synchronized using timestamp markers.

\subsection{Study Design}

\textbf{Signal processing and feature extraction.}
Physiological data were preprocessed to ensure signal integrity and cross-participant comparability. After segmenting each participant’s continuous stream into 24 trial-specific windows, we removed segments with severe artifacts and interpolated minor gaps. To normalize inter-subject variation, we applied z-score normalization within each participant.

From each trial window, we extracted handcrafted features tailored to capture time-, frequency-, and complexity-domain patterns. These included statistical descriptors (\eg, mean, skewness), spectral features (\eg, power spectral density, spectral entropy), and non-linear measures (\eg, Hurst exponent, detrended fluctuation). Feature extraction was performed separately for EDA and PPG.

To reduce dimensionality and prevent overfitting, we adopted a supervised feature selection method following \citet{pohjalainen2015feature}, using KNN-based nested cross-validation to rank features by discriminative capacity. The top-$k$ features (with $k \in [7, 15]$) were retained for model training.

\textbf{Modeling and classification tasks.}
To assess the predictive utility of physiological signals, we benchmarked four classification models: K-Nearest Neighbors (KNN), LightGBM, a simple fully connected Neural Network (NN), and a Convolutional Neural Network (CNN). This modeling suite spans from classical, interpretable methods to deep learning, enabling analysis of performance, generalizability, and robustness across modeling paradigms.

KNN served as a transparent, non-parametric baseline. Its locality-based decision mechanism proved effective for structured, low-dimensional feature sets and was also integral to the feature selection process.

LightGBM offered a powerful tree-based learner capable of capturing non-linear interactions and handling sparse or noisy input. Its regularization mechanisms and scalability made it a strong candidate for biosignal modeling.

The simple NN architecture consisted of two dense layers with ReLU activation and dropout, offering a lightweight yet expressive model to capture moderate non-linearities in the data.

CNNs were trained end-to-end on raw time-series segments, allowing temporal feature learning without manual engineering. Their inclusion tests whether biosignals contain learnable temporal patterns that handcrafted features may miss.

Each model was trained and evaluated on two tasks: binary veracity classification (True \vs~ False), and four-class joint belief-veracity classification (True-Believe, True-Not Believe, False-Believe, False-Not Believe). To isolate the contributions of each signal type, we trained models separately on EDA and PPG-derived features.

\textbf{Evaluation protocol.}
To ensure generalization to unseen users, we employed a participant-level split: 80\% of participants for training and validation (with nested cross-validation for feature selection and hyperparameter tuning), and 20\% for held-out testing. CNNs were trained on raw time windows (without feature selection), while feature-based models used the top-$k$ selected features.

This framework enables a rigorous comparison of modeling strategies, signal types, and task formulations, shedding light on when and how physiological signals encode cues relevant to truth and belief.

\section{Evaluation and Discussion}
\subsection{Evaluation}

\begin{table}[tbp]
    \centering
    \renewcommand{\arraystretch}{1.3}
    \begin{tabularx}{\linewidth}{
        l l
        >{\centering\arraybackslash}X >{\centering\arraybackslash}X >{\centering\arraybackslash}X
        >{\centering\arraybackslash}X >{\centering\arraybackslash}X >{\centering\arraybackslash}X
    }
        \toprule
        \multirow{2}{*}{\textbf{Model}} & \multirow{2}{*}{\textbf{Metric}} 
        & \multicolumn{3}{c}{\textbf{EDA}} & \multicolumn{3}{c}{\textbf{PPG}} \\
        \cmidrule(lr){3-5} \cmidrule(lr){6-8}
        & & \textbf{Macro} & \textbf{True} & \textbf{False} 
          & \textbf{Macro} & \textbf{True} & \textbf{False} \\
        \midrule

        \multirow{4}{*}{\textbf{KNN}} 
            & Accuracy   & 65.73 & --    & --    & 61.11 & --    & -- \\
            & Precision  & 65.66 & 65.93 & 65.38 & 61.11 & 65.28 & 56.94 \\
            & Recall     & 64.62 & 76.92 & 52.31 & 61.19 & 60.26 & 62.12 \\
            & F1 Score   & 64.57 & 70.92 & 58.21 & 61.04 & 62.67 & 59.41 \\
        \midrule

        \multirow{4}{*}{\textbf{LightGBM}} 
            & Accuracy   & 59.72 & --    & --    & 59.03 & --    & -- \\
            & Precision  & 64.03 & 58.06 & 70.00 & 58.52 & 60.44 & 56.60 \\
            & Recall     & 56.76 & 92.31 & 21.21 & 57.98 & 70.51 & 45.45 \\
            & F1 Score   & 51.95 & 71.20 & 32.69 & 57.75 & 65.15 & 50.34 \\
        \midrule

        \multirow{4}{*}{\textbf{NN}} 
            & Accuracy   & 57.34 & --    & --    & 59.03 & --    & -- \\
            & Precision  & 56.76 & 58.76 & 54.35 & 58.87 & 62.67 & 55.07 \\
            & Recall     & 57.34 & 73.08 & 38.46 & 58.92 & 60.26 & 57.58 \\
            & F1 Score   & 56.01 & 65.14 & 45.05 & 58.87 & 61.44 & 56.30 \\
        \midrule

        \multirow{4}{*}{\textbf{CNN}} 
            & Accuracy   & 54.48 & --    & --    & 54.07 & --    & -- \\
            & Precision  & 52.46 & 54.92 & 50.00 & 52.24 & 50.00 & 54.47 \\
            & Recall     & 50.81 & 91.78 & 9.84  & 50.73 & 9.68  & 91.78 \\
            & F1 Score   & 42.56 & 68.69 & 16.44 & 42.17 & 16.15 & 68.19 \\
        \bottomrule
    \end{tabularx}
    \caption{
        \textit{Veracity classification results using physiological features extracted from EDA and PPG signals.}
        All values are reported as percentages. For each model and signal, we report Accuracy, Precision, Recall, and F1 Score, averaged across both classes (Macro), as well as separately for the True and False classes. Dashes (--) indicate that class-specific values do not apply to Accuracy.
    }
    \label{tab:true_false_combined_all_models}
\end{table}

\textbf{Veracity classification.} Table~\ref{tab:true_false_combined_all_models} presents model performance on veracity classification using features derived from both EDA and PPG signals. Across both signal types, KNN consistently outperforms other models, achieving the highest Accuracy and Macro F1 scores. On EDA signals, KNN reaches 65.73\% Accuracy and 64.57\% Macro F1, with balanced recall across both True (76.92\%) and False (52.31\%) classes. This suggests that instance-based modeling effectively captures key discriminative patterns in EDA signals, even under moderate domain noise.

LightGBM follows closely in overall EDA performance but exhibits skewed predictions, with very high recall for True (92.31\%) and poor recall for False (21.21\%). This indicates an over-reliance on dominant class patterns, which may reflect limited sensitivity to subtle physiological indicators of falsehood. CNN, which relies on visual concept modeling \cite{plested2022deep}, performs the worst among all models on EDA, achieving only 42.56\% Macro F1 and severely underperforming on the False class (recall = 9.84\%).

On PPG features, KNN again achieves the best overall performance, with 61.11\% Accuracy and 61.04\% Macro F1. Unlike in EDA, its recall is nearly symmetric across True (60.26\%) and False (62.12\%), underscoring its robustness to the noisier and more variable nature of PPG signals. LightGBM and NN perform comparably (Macro F1 around 57–58\%), though LightGBM again over-predicts the True class. CNN shows the most extreme class imbalance on PPG, with excellent recall for False (91.78\%) but very low recall for True (9.68\%), resulting in poor overall performance (Macro F1 = 42.17\%).

Taken together, these results show that KNN is the most reliable model for binary veracity classification, especially when using EDA. Deep models like CNN struggle to generalize under the signal variability and class imbalance typical of physiological data. These findings suggest that classical models, particularly instance-based ones, may offer better robustness in low-resource, high-noise physiological settings.

\begin{table*}[tbp]
    \centering
    \renewcommand{\arraystretch}{1.3}
    \begin{tabularx}{\linewidth}{
        l l
        >{\centering\arraybackslash}X >{\centering\arraybackslash}X >{\centering\arraybackslash}X >{\centering\arraybackslash}X >{\centering\arraybackslash}X
        >{\centering\arraybackslash}X >{\centering\arraybackslash}X >{\centering\arraybackslash}X >{\centering\arraybackslash}X >{\centering\arraybackslash}X
    }
        \toprule
        \multirow{2}{*}{\textbf{Model}} & \multirow{2}{*}{\textbf{Metric}} 
        & \multicolumn{5}{c}{\textbf{EDA}} 
        & \multicolumn{5}{c}{\textbf{PPG}} \\
        \cmidrule(lr){3-7} \cmidrule(lr){8-12}
        & & \textbf{Macro} & \textbf{T-B} & \textbf{T-NB} & \textbf{F-B} & \textbf{F-NB}
          & \textbf{Macro} & \textbf{T-B} & \textbf{T-NB} & \textbf{F-B} & \textbf{F-NB} \\
        \midrule

        \multirow{4}{*}{\textbf{KNN}} 
            & Accuracy   & 37.06 & --    & --    & --    & --    & 31.94 & --    & --    & --    & -- \\
            & Precision  & 42.50 & 34.92 & 40.00 & 34.78 & 60.00 & 29.97 & 33.89 & 34.15 & 29.63 & 22.22 \\
            & Recall     & 34.75 & 48.89 & 36.36 & 44.44 & 10.34 & 30.57 & 44.44 & 42.42 & 21.62 & 13.79 \\
            & F1 Score   & 34.00 & 40.78 & 38.10 & 39.02 & 17.52 & 29.58 & 38.46 & 37.84 & 24.95 & 17.06 \\
        \midrule

        \multirow{4}{*}{\textbf{LightGBM}} 
            & Accuracy   & 33.33 & --    & --    & --    & --    & 32.64 & --    & --    & --    & -- \\
            & Precision  & 27.96 & 35.00 & 14.29 & 34.00 & 28.57 & 30.82 & 33.33 & 27.27 & 34.09 & 28.57 \\
            & Recall     & 28.75 & 60.87 & 3.13  & 43.59 & 7.41  & 29.25 & 54.35 & 9.38  & 38.46 & 14.81 \\
            & F1 Score   & 24.88 & 44.44 & 5.15  & 38.17 & 11.76 & 27.80 & 41.43 & 14.01 & 36.16 & 19.61 \\
        \midrule

        \multirow{4}{*}{\textbf{NN}} 
            & Accuracy   & 36.36 & --    & --    & --    & --    & 29.86 & --    & --    & --    & -- \\
            & Precision  & 36.99 & 44.44 & 35.71 & 35.29 & 32.50 & 27.71 & 43.90 & 17.86 & 29.09 & 20.00 \\
            & Recall     & 37.57 & 26.67 & 45.45 & 33.33 & 44.83 & 28.05 & 40.00 & 15.15 & 43.24 & 13.79 \\
            & F1 Score   & 36.33 & 33.33 & 40.00 & 34.29 & 37.68 & 27.34 & 41.86 & 16.39 & 34.78 & 16.33 \\
        \midrule

        \multirow{4}{*}{\textbf{CNN}} 
            & Accuracy   & 36.57 & --    & --    & --    & --    & 35.56 & --    & --    & --    & -- \\
            & Precision  & 38.02 & 35.43 & 50.00 & 66.67 & 0.00  & 31.44 & 36.73 & 60.00 & 29.03 & 0.00 \\
            & Recall     & 28.10 & 100.00 & 7.14 & 5.26  & 0.00  & 28.45 & 80.00 & 10.71 & 23.08 & 0.00 \\
            & F1 Score   & 18.64 & 52.29  & 12.50 & 9.76  & 0.00  & 23.56 & 50.35 & 18.18 & 25.71 & 0.00 \\
        \bottomrule
    \end{tabularx}
    \caption{
        \textit{Joint belief-veracity classification results using physiological features extracted from EDA and PPG signals.}
        All values are percentages. For each model and signal, Accuracy, Precision, Recall, and F1 Score are reported, averaged across all classes (Macro), and separately for the joint classes: T-B (True-Believe), T-NB (True-Not Believe), F-B (False-Believe), and F-NB (False-Not Believe). Dashes (--) indicate class-specific values do not apply to Accuracy.
    }
    \label{tab:joint_belief_veracity_combined}
\end{table*}

\textbf{Joint belief-veracity classification.} Table~\ref{tab:joint_belief_veracity_combined} shows model performance on the more challenging four-class joint classification task, which combines belief and veracity. Across all models and signal types, performance drops significantly compared to binary classification. This reflects the increased complexity of disentangling cognitive belief states from physiological responses, particularly when belief and veracity diverge.

KNN again achieves the highest scores on EDA, with 37.06\% Accuracy and 34.00\% Macro F1. It performs well on True-Believe (T-B) and False-Believe (F-B), but struggles with False-Not Believe (F-NB), where recall drops to 10.34\%. This suggests that disbelief in false information lacks a distinct or consistent physiological signature. LightGBM fares worse overall, with F1 scores for True-Not Believe (T-NB) and F-NB falling below 12\%, indicating difficulty in modeling less frequent or more subtle classes.

CNN shows a unique but problematic behavior: on EDA, it achieves perfect recall for T-B (100\%) but collapses on all other classes (F-NB F1 = 0.00), signaling a strong bias toward dominant patterns. On PPG, CNN achieves the highest overall accuracy (35.56\%) but again fails to identify minority classes, with F-NB F1 also at 0.00. This highlights the instability of deep temporal models under strong class imbalance and high intra-class variance.

PPG-based results exhibit similar patterns. KNN remains competitive, though its overall accuracy drops to 31.94\%. Notably, class-level F1 scores are better distributed than those of other models. NN shows slightly more balanced predictions, achieving 27.34\% Macro F1 and improved F1 for the rarest class (F-NB = 16.33\%), indicating that shallow networks may offer a trade-off between expressivity and overfitting under constrained data.

These findings underscore the inherent difficulty of modeling joint cognitive states from physiological data. While belief and veracity are cognitively distinct, their physiological signatures often overlap or remain ambiguous. Among the evaluated models, KNN shows the most consistent and interpretable performance, particularly on EDA. However, the overall low performance across all models and signal types calls for more sophisticated approaches, such as multimodal fusion \cite{wang2019loss, wang2023robust, ding2025learnable}, hierarchical modeling \cite{rouder2005introduction, kroll2010revised}, or few-shot adaptation \cite{wang20213d, wang2022uncertainty, wang2022temporal, wang2024meet}, to better capture the nuanced interplay between internal beliefs and external truth.

\subsection{Discussion}

\textbf{Physiological responses encode truth-related signals.}  
Our experimental findings demonstrate that physiological signals, particularly EDA, can capture meaningful distinctions between truthful and deceptive content. In the binary veracity classification task, models trained on EDA consistently outperformed those trained on PPG, with KNN achieving up to 65.73\% accuracy and over 64\% macro F1 score. These results lend empirical support to the hypothesis that autonomic nervous system responses, as captured through biosignals, encode latent markers of cognitive evaluation related to information veracity (see Fig.~\ref{fig:physio_overview} (a) and (b)). Notably, this effect emerges in the absence of linguistic, contextual, or behavioral cues, suggesting that bodily responses may independently reflect internal judgments of truth.

\textbf{Signal type critically affects model performance.}  
Across both binary and four-class tasks, EDA provided a more reliable signal for distinguishing cognitive states than PPG. This disparity likely stems from the neurophysiological basis of each signal type: EDA is tightly linked to sympathetic arousal and has been associated with emotional salience, conflict detection, and uncertainty, all of which may be triggered by the appraisal of potentially misleading information. In contrast, PPG captures blood volume changes, which may be influenced by a wider range of physiological and environmental factors, resulting in noisier or less class-specific patterns. Additionally, PPG-based models exhibited greater sensitivity to model choice, with CNN architectures particularly prone to overfitting, likely due to the limited data available for learning complex temporal dependencies.

\textbf{Compound cognitive constructs are harder to decode.}  
Joint belief-veracity classification, which requires distinguishing among four cognitive states (\eg, True-Believe \vs~ False-Not Believe), proved substantially more challenging than binary veracity prediction. All models struggled to distinguish the False-Not Believe condition, with F1 scores frequently near or at zero, suggesting that disbelief in misinformation may not produce a clear physiological signature, or may overlap substantially with other states such as skepticism or confusion. Even the best-performing model (KNN on EDA) achieved only 34.00\% macro F1 in this task. These findings underscore the difficulty of inferring complex, layered cognitive states solely from physiological input and point to the limitations of current modeling approaches in high-dimensional affective-cognitive classification.

\textbf{Toward physiological misinformation detection.}  
These results highlight both the promise and the constraints of using physiological signals as a foundation for misinformation detection. While binary truth discrimination appears tractable, particularly with well-engineered features and classical models, modeling belief formation, correction, or resistance requires richer input. Physiological cues may serve as a valuable, unobtrusive signal in contexts where language is inaccessible or unreliable (\eg, real-time cognitive monitoring or affective computing interfaces). However, their utility is best realized when integrated into multimodal systems that can combine biosignals with linguistic, contextual, or behavioral data to triangulate user intent and perception.

\textbf{Limitations and future directions.}  
This work has several notable limitations. First, the dataset size constrained the use of deep temporal models like CNNs and limited generalization capacity, especially in the multi-class setting. Second, we focused on two unimodal physiological signals, EDA and PPG, while omitting potentially informative modalities such as facial EMG, respiration, eye-tracking, or pupil dilation. Third, the stimuli were short factual statements presented in isolation; future work should examine richer, more emotionally or socially charged narratives to better simulate real-world information encounters. Finally, we trained and evaluated models independently on each signal type; future work should investigate joint and hierarchical fusion architectures to model cross-signal interactions, temporal synchrony, and task-specific synergies.

\textbf{Broader implications.}  
The ability to detect internal cognitive states such as belief and truth assessment via physiological signals holds promise for a range of applications, from adaptive learning systems and lie detection to human-AI collaboration and trust calibration. However, the subtlety of these states, their dependence on individual and contextual factors, and their potential ethical implications require careful methodological and normative considerations. Advancing this line of research will require not only richer data and models, but also interdisciplinary frameworks that integrate physiological computing with insights from cognitive science, communication, and social psychology.

\section{Conclusion}

This study explored whether low-cost physiological signals, electrodermal activity (EDA) and photoplethysmography (PPG), can capture how individuals respond to false information, both in terms of objective truth and subjective belief. In a controlled lab setting, we recorded participants’ physiological responses as they evaluated the veracity of climate-related claims, enabling two classification tasks: binary veracity detection and a novel four-class joint belief-veracity classification.

Our results show that EDA consistently outperformed PPG, suggesting stronger sensitivity to cognitive or affective responses associated with perceived truth. Among the evaluated models, KNN yielded the most reliable performance across both tasks. However, classification accuracy dropped substantially in the joint belief-veracity setting, underscoring the challenge of modeling the complex interplay between internal belief states and external truth conditions using biosignals alone.

These findings contribute to a growing body of work on human-centered misinformation detection by (i) providing empirical evidence that physiological signals reflect both belief and truth dimensions, (ii) introducing and evaluating a new joint classification framework, and (iii) releasing a multimodal dataset that enables further research on psychophysiological responses to misinformation.
Looking ahead, our work highlights the need for multimodal approaches that integrate physiological, cognitive, and contextual cues. Future research should explore signal fusion, personalization, and adaptive modeling to better capture the nuanced ways in which humans process, believe, and respond to misleading information.




\section{Safe and Responsible Innovation Statement}
This research explores the use of physiological signals to detect misinformation, with a strong commitment to ethical and responsible deployment. We prioritize participant privacy by using non-invasive sensors that do not capture personal identifiers. Our study actively addresses potential biases by ensuring diverse representation and considering physiological variability across different groups. To mitigate risks, we advocate for the responsible use of our findings, ensuring that they contribute solely to ethical misinformation detection systems. We are committed to transparency, inclusivity, and protecting individuals' autonomy, aiming for safe and responsible innovation in the field of multimodal interaction.

\bibliographystyle{ACM-Reference-Format}
\bibliography{bib}


\begin{thebibliography}{52}


\ifx \showCODEN    \undefined \def \showCODEN     #1{\unskip}     \fi
\ifx \showDOI      \undefined \def \showDOI       #1{#1}\fi
\ifx \showISBNx    \undefined \def \showISBNx     #1{\unskip}     \fi
\ifx \showISBNxiii \undefined \def \showISBNxiii  #1{\unskip}     \fi
\ifx \showISSN     \undefined \def \showISSN      #1{\unskip}     \fi
\ifx \showLCCN     \undefined \def \showLCCN      #1{\unskip}     \fi
\ifx \shownote     \undefined \def \shownote      #1{#1}          \fi
\ifx \showarticletitle \undefined \def \showarticletitle #1{#1}   \fi
\ifx \showURL      \undefined \def \showURL       {\relax}        \fi
\providecommand\bibfield[2]{#2}
\providecommand\bibinfo[2]{#2}
\providecommand\natexlab[1]{#1}
\providecommand\showeprint[2][]{arXiv:#2}

\bibitem[Arisoy et~al\mbox{.}(2022)]%
        {9851990}
\bibfield{author}{\bibinfo{person}{Cagri Arisoy}, \bibinfo{person}{Anuradha Mandal}, {and} \bibinfo{person}{Nitesh Saxena}.} \bibinfo{year}{2022}\natexlab{}.
\newblock \showarticletitle{Human Brains Can’t Detect Fake News: A Neuro-Cognitive Study of Textual Disinformation Susceptibility}. In \bibinfo{booktitle}{\emph{2022 19th Annual International Conference on Privacy, Security \& Trust (PST)}}. \bibinfo{pages}{1--12}.
\newblock
\urldef\tempurl%
\url{https://doi.org/10.1109/PST55820.2022.9851990}
\showDOI{\tempurl}


\bibitem[Babaei et~al\mbox{.}(2021)]%
        {babaei2021critique}
\bibfield{author}{\bibinfo{person}{Ebrahim Babaei}, \bibinfo{person}{Benjamin Tag}, \bibinfo{person}{Tilman Dingler}, {and} \bibinfo{person}{Eduardo Velloso}.} \bibinfo{year}{2021}\natexlab{}.
\newblock \showarticletitle{A critique of electrodermal activity practices at chi}. In \bibinfo{booktitle}{\emph{Proceedings of the 2021 CHI Conference on Human Factors in Computing Systems}}. \bibinfo{pages}{1--14}.
\newblock


\bibitem[Bhaskaran et~al\mbox{.}(2011)]%
        {bhaskaran2011lie}
\bibfield{author}{\bibinfo{person}{Nisha Bhaskaran}, \bibinfo{person}{Ifeoma Nwogu}, \bibinfo{person}{Mark~G Frank}, {and} \bibinfo{person}{Venu Govindaraju}.} \bibinfo{year}{2011}\natexlab{}.
\newblock \showarticletitle{Lie to me: Deceit detection via online behavioral learning}. In \bibinfo{booktitle}{\emph{2011 IEEE International Conference on Automatic Face \& Gesture Recognition (FG)}}. IEEE, \bibinfo{pages}{24--29}.
\newblock


\bibitem[Bhattacharjee et~al\mbox{.}(2017)]%
        {bhattacharjee2017active}
\bibfield{author}{\bibinfo{person}{Sreyasee~Das Bhattacharjee}, \bibinfo{person}{Ashit Talukder}, {and} \bibinfo{person}{Bala~Venkatram Balantrapu}.} \bibinfo{year}{2017}\natexlab{}.
\newblock \showarticletitle{Active learning based news veracity detection with feature weighting and deep-shallow fusion}. In \bibinfo{booktitle}{\emph{2017 IEEE International Conference on Big Data (Big Data)}}. IEEE, \bibinfo{pages}{556--565}.
\newblock


\bibitem[Chen et~al\mbox{.}(2024)]%
        {chen2024sato}
\bibfield{author}{\bibinfo{person}{Wenshuo Chen}, \bibinfo{person}{Hongru Xiao}, \bibinfo{person}{Erhang Zhang}, \bibinfo{person}{Lijie Hu}, \bibinfo{person}{Lei Wang}, \bibinfo{person}{Mengyuan Liu}, {and} \bibinfo{person}{Chen Chen}.} \bibinfo{year}{2024}\natexlab{}.
\newblock \showarticletitle{SATO: Stable Text-to-Motion Framework}. In \bibinfo{booktitle}{\emph{Proceedings of the 32nd ACM International Conference on Multimedia}}. \bibinfo{pages}{6989--6997}.
\newblock


\bibitem[Chu et~al\mbox{.}(2021)]%
        {chu2021detecting}
\bibfield{author}{\bibinfo{person}{Ruimin Chu}, \bibinfo{person}{Jessica~Sharmin Rahman}, \bibinfo{person}{Sabrina Caldwell}, \bibinfo{person}{Xuanying Zhu}, {and} \bibinfo{person}{Tom Gedeon}.} \bibinfo{year}{2021}\natexlab{}.
\newblock \showarticletitle{Detecting Lies: Finding the Degree of Falsehood from Observers’ Physiological Responses}. In \bibinfo{booktitle}{\emph{2021 IEEE International Conference on Systems, Man, and Cybernetics (SMC)}}. IEEE, \bibinfo{pages}{1959--1965}.
\newblock


\bibitem[Council et~al\mbox{.}(2003)]%
        {national2003polygraph}
\bibfield{author}{\bibinfo{person}{National~Research Council}, \bibinfo{person}{Division of Behavioral}, \bibinfo{person}{Committee on National~Statistics}, \bibinfo{person}{Board on Behavioral}, \bibinfo{person}{Sensory Sciences}, {and} \bibinfo{person}{Committee to~Review the Scientific Evidence on~the Polygraph}.} \bibinfo{year}{2003}\natexlab{}.
\newblock \bibinfo{booktitle}{\emph{The polygraph and lie detection}}.
\newblock \bibinfo{publisher}{National Academies Press}.
\newblock


\bibitem[Datta et~al\mbox{.}(2024)]%
        {datta2024enhancing}
\bibfield{author}{\bibinfo{person}{KS~Sreekar Datta}, \bibinfo{person}{G~Narasimha Naidu}, \bibinfo{person}{S Abhishek}, {et~al\mbox{.}}} \bibinfo{year}{2024}\natexlab{}.
\newblock \showarticletitle{Enhancing Veracity: Empirical Evaluation of Fake News Detection Techniques}.
\newblock \bibinfo{journal}{\emph{Procedia Computer Science}}  \bibinfo{volume}{233} (\bibinfo{year}{2024}), \bibinfo{pages}{97--107}.
\newblock


\bibitem[De~Filippi et~al\mbox{.}({[n.\,d.]})]%
        {deseeing}
\bibfield{author}{\bibinfo{person}{Eleonora De~Filippi}, \bibinfo{person}{Arijit Nandi}, \bibinfo{person}{Julian Vicens}, \bibinfo{person}{Elena Alvarez-Garc{\i}a}, \bibinfo{person}{Daniel Garc{\i}a-Costa}, \bibinfo{person}{Francisco Grimaldo}, {and} \bibinfo{person}{Alexandre Pereda-Banos}.} \bibinfo{year}{[n.\,d.]}\natexlab{}.
\newblock \showarticletitle{Seeing through lies: Correlation between physiological and sentiment metrics in the consumption and dissemination of fake news}.
\newblock  (\bibinfo{year}{[n.\,d.]}).
\newblock


\bibitem[Ding et~al\mbox{.}(2025)]%
        {ding2025learnable}
\bibfield{author}{\bibinfo{person}{Dexuan Ding}, \bibinfo{person}{Lei Wang}, \bibinfo{person}{Liyun Zhu}, \bibinfo{person}{Tom Gedeon}, {and} \bibinfo{person}{Piotr Koniusz}.} \bibinfo{year}{2025}\natexlab{}.
\newblock \showarticletitle{Learnable Expansion of Graph Operators for Multi-Modal Feature Fusion}. In \bibinfo{booktitle}{\emph{The Thirteenth International Conference on Learning Representations}}.
\newblock
\urldef\tempurl%
\url{https://openreview.net/forum?id=SMZqIOSdlN}
\showURL{%
\tempurl}


\bibitem[Ding and Wang(2025a)]%
        {10.1145/3701716.3717744}
\bibfield{author}{\bibinfo{person}{Xi Ding} {and} \bibinfo{person}{Lei Wang}.} \bibinfo{year}{2025}\natexlab{a}.
\newblock \showarticletitle{Do Language Models Understand Time?}. In \bibinfo{booktitle}{\emph{Companion Proceedings of the ACM Web Conference 2025}} (Sydney, NSW, Australia) \emph{(\bibinfo{series}{WWW '25 Companion})}. \bibinfo{publisher}{Association for Computing Machinery}, \bibinfo{address}{New York, NY, USA}.
\newblock
\showISBNx{9798400713316}
\urldef\tempurl%
\url{https://doi.org/10.1145/3701716.3717744}
\showDOI{\tempurl}


\bibitem[Ding and Wang(2025b)]%
        {dingjourney}
\bibfield{author}{\bibinfo{person}{Xi Ding} {and} \bibinfo{person}{Lei Wang}.} \bibinfo{year}{2025}\natexlab{b}.
\newblock \showarticletitle{The Journey of Action Recognition}. In \bibinfo{booktitle}{\emph{Companion Proceedings of the ACM Web Conference 2025}} (Sydney, NSW, Australia) \emph{(\bibinfo{series}{WWW '25 Companion})}. \bibinfo{publisher}{Association for Computing Machinery}, \bibinfo{address}{New York, NY, USA}.
\newblock
\showISBNx{9798400713316}
\urldef\tempurl%
\url{https://doi.org/10.1145/3701716.3717746}
\showDOI{\tempurl}


\bibitem[Fang et~al\mbox{.}(2024)]%
        {fang2024signllm}
\bibfield{author}{\bibinfo{person}{Sen Fang}, \bibinfo{person}{Lei Wang}, \bibinfo{person}{Ce Zheng}, \bibinfo{person}{Yapeng Tian}, {and} \bibinfo{person}{Chen Chen}.} \bibinfo{year}{2024}\natexlab{}.
\newblock \showarticletitle{Signllm: Sign languages production large language models}.
\newblock \bibinfo{journal}{\emph{arXiv preprint arXiv:2405.10718}} (\bibinfo{year}{2024}).
\newblock


\bibitem[Gasparini et~al\mbox{.}(2021)]%
        {gasparini2021deep}
\bibfield{author}{\bibinfo{person}{Francesca Gasparini}, \bibinfo{person}{Alessandra Grossi}, {and} \bibinfo{person}{Stefania Bandini}.} \bibinfo{year}{2021}\natexlab{}.
\newblock \showarticletitle{A deep learning approach to recognize cognitive load using ppg signals}. In \bibinfo{booktitle}{\emph{Proceedings of the 14th PErvasive technologies related to assistive environments conference}}. \bibinfo{pages}{489--495}.
\newblock


\bibitem[Gunderson et~al\mbox{.}(2023)]%
        {gunderson2023body}
\bibfield{author}{\bibinfo{person}{Christopher~A Gunderson}, \bibinfo{person}{LM ten Brinke}, {and} \bibinfo{person}{Peter Sokol-Hessner}.} \bibinfo{year}{2023}\natexlab{}.
\newblock \showarticletitle{When the body knows: Interoceptive accuracy enhances physiological but not explicit differentiation between liars and truth-tellers}.
\newblock \bibinfo{journal}{\emph{Personality and Individual Differences}}  \bibinfo{volume}{204} (\bibinfo{year}{2023}), \bibinfo{pages}{112039}.
\newblock


\bibitem[Hartwig and Bond~Jr(2014)]%
        {hartwig2014lie}
\bibfield{author}{\bibinfo{person}{Maria Hartwig} {and} \bibinfo{person}{Charles~F Bond~Jr}.} \bibinfo{year}{2014}\natexlab{}.
\newblock \showarticletitle{Lie detection from multiple cues: A meta-analysis}.
\newblock \bibinfo{journal}{\emph{Applied Cognitive Psychology}} \bibinfo{volume}{28}, \bibinfo{number}{5} (\bibinfo{year}{2014}), \bibinfo{pages}{661--676}.
\newblock


\bibitem[Horvath(1982)]%
        {horvath1982detecting}
\bibfield{author}{\bibinfo{person}{Frank Horvath}.} \bibinfo{year}{1982}\natexlab{}.
\newblock \showarticletitle{Detecting deception: the promise and the reality of voice stress analysis}.
\newblock \bibinfo{journal}{\emph{Journal of Forensic Sciences}} \bibinfo{volume}{27}, \bibinfo{number}{2} (\bibinfo{year}{1982}), \bibinfo{pages}{340--351}.
\newblock


\bibitem[Jiang et~al\mbox{.}(2024)]%
        {jiang2024repetition}
\bibfield{author}{\bibinfo{person}{Yangxueqing Jiang}, \bibinfo{person}{Norbert Schwarz}, \bibinfo{person}{Katherine~J Reynolds}, {and} \bibinfo{person}{Eryn~J Newman}.} \bibinfo{year}{2024}\natexlab{}.
\newblock \showarticletitle{Repetition increases belief in climate-skeptical claims, even for climate science endorsers}.
\newblock \bibinfo{journal}{\emph{Plos one}} \bibinfo{volume}{19}, \bibinfo{number}{8} (\bibinfo{year}{2024}), \bibinfo{pages}{e0307294}.
\newblock


\bibitem[Joshi et~al\mbox{.}(2025)]%
        {joshi2025multimodal}
\bibfield{author}{\bibinfo{person}{Gargi Joshi}, \bibinfo{person}{Vaibhav Tasgaonkar}, \bibinfo{person}{Aditya Deshpande}, \bibinfo{person}{Aditya Desai}, \bibinfo{person}{Bhavya Shah}, \bibinfo{person}{Akshay Kushawaha}, \bibinfo{person}{Aadith Sukumar}, \bibinfo{person}{Kermi Kotecha}, \bibinfo{person}{Saumit Kunder}, \bibinfo{person}{Yoginii Waykole}, {et~al\mbox{.}}} \bibinfo{year}{2025}\natexlab{}.
\newblock \showarticletitle{Multimodal machine learning for deception detection using behavioral and physiological data}.
\newblock \bibinfo{journal}{\emph{Scientific Reports}} \bibinfo{volume}{15}, \bibinfo{number}{1} (\bibinfo{year}{2025}), \bibinfo{pages}{8943}.
\newblock


\bibitem[Kang and Kim(2022)]%
        {kang20221d}
\bibfield{author}{\bibinfo{person}{Dong-Hyun Kang} {and} \bibinfo{person}{Deok-Hwan Kim}.} \bibinfo{year}{2022}\natexlab{}.
\newblock \showarticletitle{1D convolutional autoencoder-based PPG and GSR signals for real-time emotion classification}.
\newblock \bibinfo{journal}{\emph{IEEE Access}}  \bibinfo{volume}{10} (\bibinfo{year}{2022}), \bibinfo{pages}{91332--91345}.
\newblock


\bibitem[Kroll et~al\mbox{.}(2010)]%
        {kroll2010revised}
\bibfield{author}{\bibinfo{person}{Judith~F Kroll}, \bibinfo{person}{Janet~G Van~Hell}, \bibinfo{person}{Natasha Tokowicz}, {and} \bibinfo{person}{David~W Green}.} \bibinfo{year}{2010}\natexlab{}.
\newblock \showarticletitle{The Revised Hierarchical Model: A critical review and assessment}.
\newblock \bibinfo{journal}{\emph{Bilingualism: Language and Cognition}} \bibinfo{volume}{13}, \bibinfo{number}{3} (\bibinfo{year}{2010}), \bibinfo{pages}{373--381}.
\newblock


\bibitem[Kumar et~al\mbox{.}(2019)]%
        {kumar2019rumour}
\bibfield{author}{\bibinfo{person}{Akshi Kumar}, \bibinfo{person}{Saurabh~Raj Sangwan}, {and} \bibinfo{person}{Anand Nayyar}.} \bibinfo{year}{2019}\natexlab{}.
\newblock \showarticletitle{Rumour veracity detection on twitter using particle swarm optimized shallow classifiers}.
\newblock \bibinfo{journal}{\emph{Multimedia Tools and Applications}}  \bibinfo{volume}{78} (\bibinfo{year}{2019}), \bibinfo{pages}{24083--24101}.
\newblock


\bibitem[Lin and Li(2023)]%
        {lin2023review}
\bibfield{author}{\bibinfo{person}{Wenqian Lin} {and} \bibinfo{person}{Chao Li}.} \bibinfo{year}{2023}\natexlab{}.
\newblock \showarticletitle{Review of studies on emotion recognition and judgment based on physiological signals}.
\newblock \bibinfo{journal}{\emph{Applied Sciences}} \bibinfo{volume}{13}, \bibinfo{number}{4} (\bibinfo{year}{2023}), \bibinfo{pages}{2573}.
\newblock


\bibitem[Molina et~al\mbox{.}(2025)]%
        {molina2025exploring}
\bibfield{author}{\bibinfo{person}{R Molina}, \bibinfo{person}{Y Crespo}, \bibinfo{person}{JR {\'A}rbol}, \bibinfo{person}{AV Arias-Ordu{\~n}a}, \bibinfo{person}{AJ Ib{\'a}{\~n}ez-Molina}, {and} \bibinfo{person}{S Iglesias-Parro}.} \bibinfo{year}{2025}\natexlab{}.
\newblock \showarticletitle{Exploring the Neurophysiological Basis of Misinformation: A Behavioral and Neural Complexity Analysis}.
\newblock \bibinfo{journal}{\emph{Behavioural Brain Research}} (\bibinfo{year}{2025}), \bibinfo{pages}{115592}.
\newblock


\bibitem[Morozova et~al\mbox{.}(2024)]%
        {10705859}
\bibfield{author}{\bibinfo{person}{Alexandra Morozova}, \bibinfo{person}{Eliana Monahhova}, \bibinfo{person}{Julia Gorodnicheva}, \bibinfo{person}{Oksana Zinchenko}, \bibinfo{person}{Anna Shestakova}, {and} \bibinfo{person}{Vasily Klucharev}.} \bibinfo{year}{2024}\natexlab{}.
\newblock \showarticletitle{Event-Related Potentials in Response to Fake News Correction: Pilot Study}. In \bibinfo{booktitle}{\emph{2024 Sixth International Conference Neurotechnologies and Neurointerfaces (CNN)}}. \bibinfo{pages}{124--127}.
\newblock
\urldef\tempurl%
\url{https://doi.org/10.1109/CNN63506.2024.10705859}
\showDOI{\tempurl}


\bibitem[Pandey et~al\mbox{.}(2019)]%
        {pandey2019techniques}
\bibfield{author}{\bibinfo{person}{Payal Pandey}, \bibinfo{person}{Divyansh Thakur}, {and} \bibinfo{person}{Bishan Thakur}.} \bibinfo{year}{2019}\natexlab{}.
\newblock \showarticletitle{Techniques for Behavior Lie Detection with the Aid of Physiological Signals: A Review}. In \bibinfo{booktitle}{\emph{Proceedings of the International Conference on Advances in Electronics, Electrical \& Computational Intelligence (ICAEEC)}}.
\newblock


\bibitem[Plested and Gedeon(2022)]%
        {plested2022deep}
\bibfield{author}{\bibinfo{person}{Jo Plested} {and} \bibinfo{person}{Tom Gedeon}.} \bibinfo{year}{2022}\natexlab{}.
\newblock \showarticletitle{Deep transfer learning for image classification: a survey}.
\newblock \bibinfo{journal}{\emph{arXiv preprint arXiv:2205.09904}} (\bibinfo{year}{2022}).
\newblock


\bibitem[Pohjalainen et~al\mbox{.}(2015)]%
        {pohjalainen2015feature}
\bibfield{author}{\bibinfo{person}{Jouni Pohjalainen}, \bibinfo{person}{Okko R{\"a}s{\"a}nen}, {and} \bibinfo{person}{Serdar Kadioglu}.} \bibinfo{year}{2015}\natexlab{}.
\newblock \showarticletitle{Feature selection methods and their combinations in high-dimensional classification of speaker likability, intelligibility and personality traits}.
\newblock \bibinfo{journal}{\emph{Computer Speech \& Language}} \bibinfo{volume}{29}, \bibinfo{number}{1} (\bibinfo{year}{2015}), \bibinfo{pages}{145--171}.
\newblock


\bibitem[Rahal et~al\mbox{.}(2024)]%
        {rahal2024your}
\bibfield{author}{\bibinfo{person}{Rima-Maria Rahal}, \bibinfo{person}{Teun Siebers}, \bibinfo{person}{Willem~WA Sleegers}, {and} \bibinfo{person}{Ilja van Beest}.} \bibinfo{year}{2024}\natexlab{}.
\newblock \showarticletitle{Your lies don't leave me cold: Assessing direct, indirect and physiological measures of lie detection}.
\newblock \bibinfo{journal}{\emph{Acta Psychologica}}  \bibinfo{volume}{251} (\bibinfo{year}{2024}), \bibinfo{pages}{104548}.
\newblock


\bibitem[Rouder and Lu(2005)]%
        {rouder2005introduction}
\bibfield{author}{\bibinfo{person}{Jeffrey~N Rouder} {and} \bibinfo{person}{Jun Lu}.} \bibinfo{year}{2005}\natexlab{}.
\newblock \showarticletitle{An introduction to Bayesian hierarchical models with an application in the theory of signal detection}.
\newblock \bibinfo{journal}{\emph{Psychonomic bulletin \& review}} \bibinfo{volume}{12}, \bibinfo{number}{4} (\bibinfo{year}{2005}), \bibinfo{pages}{573--604}.
\newblock


\bibitem[Speth et~al\mbox{.}(2021)]%
        {speth2021deception}
\bibfield{author}{\bibinfo{person}{Jeremy Speth}, \bibinfo{person}{Nathan Vance}, \bibinfo{person}{Adam Czajka}, \bibinfo{person}{Kevin~W Bowyer}, \bibinfo{person}{Diane Wright}, {and} \bibinfo{person}{Patrick Flynn}.} \bibinfo{year}{2021}\natexlab{}.
\newblock \showarticletitle{Deception detection and remote physiological monitoring: A dataset and baseline experimental results}. In \bibinfo{booktitle}{\emph{2021 IEEE International Joint Conference on Biometrics (IJCB)}}. IEEE, \bibinfo{pages}{1--8}.
\newblock


\bibitem[Synnott et~al\mbox{.}(2016)]%
        {synnott2016review}
\bibfield{author}{\bibinfo{person}{John Synnott}, \bibinfo{person}{Maria Ioannou}, {and} \bibinfo{person}{Anita Fumagalli}.} \bibinfo{year}{2016}\natexlab{}.
\newblock \showarticletitle{A review of the polygraph: history, current status and emerging research}.
\newblock \bibinfo{journal}{\emph{Custodial Review}} (\bibinfo{year}{2016}), \bibinfo{pages}{22--23}.
\newblock


\bibitem[Talaat(2024)]%
        {talaat2024explainable}
\bibfield{author}{\bibinfo{person}{Fatma~M Talaat}.} \bibinfo{year}{2024}\natexlab{}.
\newblock \showarticletitle{Explainable enhanced recurrent neural network for lie detection using voice stress analysis}.
\newblock \bibinfo{journal}{\emph{Multimedia Tools and Applications}} \bibinfo{volume}{83}, \bibinfo{number}{11} (\bibinfo{year}{2024}), \bibinfo{pages}{32277--32299}.
\newblock


\bibitem[Ten~Brinke et~al\mbox{.}(2019)]%
        {ten2019different}
\bibfield{author}{\bibinfo{person}{Leanne Ten~Brinke}, \bibinfo{person}{Julia~J Lee}, {and} \bibinfo{person}{Dana~R Carney}.} \bibinfo{year}{2019}\natexlab{}.
\newblock \showarticletitle{Different physiological reactions when observing lies versus truths: Initial evidence and an intervention to enhance accuracy.}
\newblock \bibinfo{journal}{\emph{Journal of personality and social psychology}} \bibinfo{volume}{117}, \bibinfo{number}{3} (\bibinfo{year}{2019}), \bibinfo{pages}{560}.
\newblock


\bibitem[Ulman({[n.\,d.]})]%
        {ulmandevelopment}
\bibfield{author}{\bibinfo{person}{Hana~K Ulman}.} \bibinfo{year}{[n.\,d.]}\natexlab{}.
\newblock \showarticletitle{Development of a Physiological Monitoring System for Credibility Assessment}.
\newblock  (\bibinfo{year}{[n.\,d.]}).
\newblock


\bibitem[van’t Veer et~al\mbox{.}(2015)]%
        {van2015unconscious}
\bibfield{author}{\bibinfo{person}{Anna~E van’t Veer}, \bibinfo{person}{Marcello Gallucci}, \bibinfo{person}{Mari{\"e}lle Stel}, {and} \bibinfo{person}{Ilja~van Beest}.} \bibinfo{year}{2015}\natexlab{}.
\newblock \showarticletitle{Unconscious deception detection measured by finger skin temperature and indirect veracity judgments—results of a registered report}.
\newblock \bibinfo{journal}{\emph{Frontiers in psychology}}  \bibinfo{volume}{6} (\bibinfo{year}{2015}), \bibinfo{pages}{672}.
\newblock


\bibitem[Volz et~al\mbox{.}(2017)]%
        {volz2017psychophysiological}
\bibfield{author}{\bibinfo{person}{Katja Volz}, \bibinfo{person}{Rainer Leonhart}, \bibinfo{person}{Rudolf Stark}, \bibinfo{person}{Dieter Vaitl}, {and} \bibinfo{person}{Wolfgang Ambach}.} \bibinfo{year}{2017}\natexlab{}.
\newblock \showarticletitle{Psychophysiological correlates of the misinformation effect}.
\newblock \bibinfo{journal}{\emph{International Journal of Psychophysiology}}  \bibinfo{volume}{117} (\bibinfo{year}{2017}), \bibinfo{pages}{1--9}.
\newblock


\bibitem[Wahab({[n.\,d.]})]%
        {wahabassessing}
\bibfield{author}{\bibinfo{person}{Md~Imran Wahab}.} \bibinfo{year}{[n.\,d.]}\natexlab{}.
\newblock \showarticletitle{Assessing the Reliability and Evidentiary Value of Polygraph Tests}.
\newblock  (\bibinfo{year}{[n.\,d.]}).
\newblock


\bibitem[Wang(2017)]%
        {wang2017analysis}
\bibfield{author}{\bibinfo{person}{Lei Wang}.} \bibinfo{year}{2017}\natexlab{}.
\newblock \emph{\bibinfo{title}{Analysis and Evaluation of {K}inect-based Action Recognition Algorithms}}.
\newblock \bibinfo{thesistype}{Master's\ thesis}. \bibinfo{school}{School of the Computer Science and Software Engineering, The University of Western Australia}.
\newblock


\bibitem[Wang(2023)]%
        {wang2023robust}
\bibfield{author}{\bibinfo{person}{Lei Wang}.} \bibinfo{year}{2023}\natexlab{}.
\newblock \emph{\bibinfo{title}{Robust human action modelling}}.
\newblock \bibinfo{thesistype}{Ph.\,D. Dissertation}. \bibinfo{school}{The Australian National University (Australia)}.
\newblock


\bibitem[Wang et~al\mbox{.}(2025a)]%
        {wang2025time}
\bibfield{author}{\bibinfo{person}{Lei Wang}, \bibinfo{person}{Md~Zakir Hossain}, \bibinfo{person}{Syed M.~S. Islam}, \bibinfo{person}{Tom Gedeon}, \bibinfo{person}{Sharifa Alghowinem}, \bibinfo{person}{Isabella Yu}, \bibinfo{person}{Serena Bono}, \bibinfo{person}{Xuanying Zhu}, \bibinfo{person}{Gennie Nguyen}, \bibinfo{person}{Nur Al~Hasan Haldar}, \bibinfo{person}{Seyed Mohammad~Jafar Jalali}, \bibinfo{person}{Md~Abdur Razzaque}, \bibinfo{person}{Imran Razzak}, \bibinfo{person}{Rafiqul Islam}, \bibinfo{person}{Shahadat Uddin}, \bibinfo{person}{Naeem~Khalid Janjua}, \bibinfo{person}{Aneesh Krishna}, {and} \bibinfo{person}{Manzur Ashraf}.} \bibinfo{year}{2025}\natexlab{a}.
\newblock \showarticletitle{{TIME} 2025: 1st International Workshop on Transformative Insights in Multi-faceted Evaluation}. In \bibinfo{booktitle}{\emph{The First International Workshop on Transformative Insights in Multifaceted Evaluation at The Web Conference 2025}}.
\newblock
\urldef\tempurl%
\url{https://openreview.net/forum?id=Ns2pk0ePyG}
\showURL{%
\tempurl}


\bibitem[Wang et~al\mbox{.}(2019)]%
        {wang2019loss}
\bibfield{author}{\bibinfo{person}{Lei Wang}, \bibinfo{person}{Du~Q Huynh}, {and} \bibinfo{person}{Moussa~Reda Mansour}.} \bibinfo{year}{2019}\natexlab{}.
\newblock \showarticletitle{Loss switching fusion with similarity search for video classification}. In \bibinfo{booktitle}{\emph{2019 IEEE international conference on image processing (ICIP)}}. IEEE, \bibinfo{pages}{974--978}.
\newblock


\bibitem[Wang and Koniusz(2022a)]%
        {wang2022temporal}
\bibfield{author}{\bibinfo{person}{Lei Wang} {and} \bibinfo{person}{Piotr Koniusz}.} \bibinfo{year}{2022}\natexlab{a}.
\newblock \showarticletitle{Temporal-viewpoint transportation plan for skeletal few-shot action recognition}. In \bibinfo{booktitle}{\emph{Proceedings of the Asian Conference on Computer Vision}}. \bibinfo{pages}{4176--4193}.
\newblock


\bibitem[Wang and Koniusz(2022b)]%
        {wang2022uncertainty}
\bibfield{author}{\bibinfo{person}{Lei Wang} {and} \bibinfo{person}{Piotr Koniusz}.} \bibinfo{year}{2022}\natexlab{b}.
\newblock \showarticletitle{Uncertainty-dtw for time series and sequences}. In \bibinfo{booktitle}{\emph{European Conference on Computer Vision}}. Springer, \bibinfo{pages}{176--195}.
\newblock


\bibitem[Wang and Koniusz(2024)]%
        {wang2024flow}
\bibfield{author}{\bibinfo{person}{Lei Wang} {and} \bibinfo{person}{Piotr Koniusz}.} \bibinfo{year}{2024}\natexlab{}.
\newblock \showarticletitle{Flow dynamics correction for action recognition}. In \bibinfo{booktitle}{\emph{ICASSP 2024-2024 IEEE International Conference on Acoustics, Speech and Signal Processing (ICASSP)}}. IEEE, \bibinfo{pages}{3795--3799}.
\newblock


\bibitem[Wang et~al\mbox{.}(2021)]%
        {wang20213d}
\bibfield{author}{\bibinfo{person}{Lei Wang}, \bibinfo{person}{Jun Liu}, {and} \bibinfo{person}{Piotr Koniusz}.} \bibinfo{year}{2021}\natexlab{}.
\newblock \showarticletitle{3D Skeleton-based Few-shot Action Recognition with JEANIE is not so Na\"ive}.
\newblock \bibinfo{journal}{\emph{arXiv preprint arXiv:2112.12668}} (\bibinfo{year}{2021}).
\newblock


\bibitem[Wang et~al\mbox{.}(2024a)]%
        {wang2024meet}
\bibfield{author}{\bibinfo{person}{Lei Wang}, \bibinfo{person}{Jun Liu}, \bibinfo{person}{Liang Zheng}, \bibinfo{person}{Tom Gedeon}, {and} \bibinfo{person}{Piotr Koniusz}.} \bibinfo{year}{2024}\natexlab{a}.
\newblock \showarticletitle{Meet jeanie: a similarity measure for 3d skeleton sequences via temporal-viewpoint alignment}.
\newblock \bibinfo{journal}{\emph{International Journal of Computer Vision}} \bibinfo{volume}{132}, \bibinfo{number}{9} (\bibinfo{year}{2024}), \bibinfo{pages}{4091--4122}.
\newblock


\bibitem[Wang et~al\mbox{.}(2024b)]%
        {wang2024high}
\bibfield{author}{\bibinfo{person}{Lei Wang}, \bibinfo{person}{Ke Sun}, {and} \bibinfo{person}{Piotr Koniusz}.} \bibinfo{year}{2024}\natexlab{b}.
\newblock \showarticletitle{High-order tensor pooling with attention for action recognition}. In \bibinfo{booktitle}{\emph{ICASSP 2024-2024 IEEE International Conference on Acoustics, Speech and Signal Processing (ICASSP)}}. IEEE, \bibinfo{pages}{3885--3889}.
\newblock


\bibitem[Wang et~al\mbox{.}({[n.\,d.]})]%
        {wangtaylor}
\bibfield{author}{\bibinfo{person}{Lei Wang}, \bibinfo{person}{Xiuyuan Yuan}, \bibinfo{person}{Tom Gedeon}, {and} \bibinfo{person}{Liang Zheng}.} \bibinfo{year}{[n.\,d.]}\natexlab{}.
\newblock \showarticletitle{Taylor Videos for Action Recognition}. In \bibinfo{booktitle}{\emph{Forty-first International Conference on Machine Learning}}.
\newblock


\bibitem[Wang et~al\mbox{.}(2025b)]%
        {wang2025bridging}
\bibfield{author}{\bibinfo{person}{Zihan Wang}, \bibinfo{person}{Lu Yuan}, \bibinfo{person}{Zhengxuan Zhang}, {and} \bibinfo{person}{Qing Zhao}.} \bibinfo{year}{2025}\natexlab{b}.
\newblock \showarticletitle{Bridging Cognition and Emotion: Empathy-Driven Multimodal Misinformation Detection}.
\newblock \bibinfo{journal}{\emph{arXiv preprint arXiv:2504.17332}} (\bibinfo{year}{2025}).
\newblock


\bibitem[Zhu et~al\mbox{.}({[n.\,d.]})]%
        {zhuadvancing}
\bibfield{author}{\bibinfo{person}{Liyun Zhu}, \bibinfo{person}{Lei Wang}, \bibinfo{person}{Arjun Raj}, \bibinfo{person}{Tom Gedeon}, {and} \bibinfo{person}{Chen Chen}.} \bibinfo{year}{[n.\,d.]}\natexlab{}.
\newblock \showarticletitle{Advancing Video Anomaly Detection: A Concise Review and a New Dataset}. In \bibinfo{booktitle}{\emph{The Thirty-eight Conference on Neural Information Processing Systems Datasets and Benchmarks Track}}.
\newblock


\bibitem[Zhu et~al\mbox{.}(2018)]%
        {zhu2018detecting}
\bibfield{author}{\bibinfo{person}{Xuanying Zhu}, \bibinfo{person}{Zhenyue Qin}, \bibinfo{person}{Tom Gedeon}, \bibinfo{person}{Richard Jones}, \bibinfo{person}{Md~Zakir Hossain}, {and} \bibinfo{person}{Sabrina Caldwell}.} \bibinfo{year}{2018}\natexlab{}.
\newblock \showarticletitle{Detecting the doubt effect and subjective beliefs using neural networks and observers’ pupillary responses}. In \bibinfo{booktitle}{\emph{Neural Information Processing: 25th International Conference, ICONIP 2018, Siem Reap, Cambodia, December 13-16, 2018, Proceedings, Part IV 25}}. Springer, \bibinfo{pages}{610--621}.
\newblock


\end{thebibliography}










\end{document}